# Non-Diffracting Electron Vortex Beams Balancing Their Electron-Electron Interactions

Maor Mutzafi[1], Ido Kaminer[2,1], Gal Harari[1] and Mordechai Segev[1]

1   Physics Department and Solid State Institute, Technion, Haifa 32000, Israel

2   Department of Physics, Massachusetts Institute of Technology, 77 Massachusetts Avenue, Cambridge, Massachusetts, USA

**Abstract**

By introducing concepts of beam shaping into quantum mechanics, we show how interference effects of the quantum wavefunction describing multiple electrons can exactly balance the repulsion among the electrons. With proper shaping of the electron wavefunction, we propose non-diffracting quantum wavepackets of multiple electrons that can also carry orbital angular momentum, in the form of multi-electron non-diffracting vortex beams. The wavefunction is designed to compensate for both the repulsion between electrons and for the diffraction-broadening. This wavefunction shaping facilitates the use of electron beams of higher current in numerous applications, thereby improving the signal-to-noise-ratio in electron microscopy and related systems without compromising on the spatial resolution. Our scheme potentially applies for any beams of charged particles, such as protons, muons and ion beams.

The wave-like nature of electrons is now a well-established concept for many years, with famous experimental demonstrations such as the double slit experiment[1] and Snell's like refraction[2]. The wavelength of an electron with accessible energy is several orders of magnitude shorter than optical wavelengths, thereby providing access to microscopy experiments at atomic resolution[3]. Naturally, electron beam sources find numerous applications beyond microscopy, including radiation sources such as free electron lasers[4,5], electron beam lithography, etc. Yet only in the past several years, the actual shaping of the wavefunction of electrons has become experimentally possible[6]. Indeed, shaping and manipulating the wavefunction of an electron is currently achieved through new techniques that use binary masks[7] (made from a thin metal foil fabricated at nano-scale resolution) or amplitude and phase masks[8] (made of thin Silicon-Nitride (SiN) membranes) imprinting the actual amplitude and phase distribution of the wavefunction. With these techniques, shaping the quantum wavefunction of electron beams (EBeams) has been used to generate EBeams carrying orbital angular momentum (OAM)[6,7,9,10] self-accelerating (Airy) EBeams[11] and more recently - Bessel Ebeams[12]. Such techniques may fundamentally change all EBeam applications and experiments, since they allow direct control over the quantum wavefunction of electrons[13].

One of the most important applications of EBeams is electron microscopy, which has become an essential tool in many fields of science and technology, such as biology, materials science, electrical engineering and more. Scanning Electron Microscopy[14] (SEM), Transmission Electron Microscopy[15] (TEM) and Scanning Transmission Electron Microscopy (STEM) produce images by scanning a sample with a focused EBeam or transmitting an EBeam through the sample. The EBeam interacts with the sample and produces an image containing information that is often at atomic resolution. Importantly, the fundamental limit on the highest resolution possible in electron microscopy is the wavelength of the particle, which for electrons is on the order of pico-meters ($10^{-12}$m). In practice, however, state-of-the-art electron microscopes are still 2-3 orders of magnitude away from this fundamental limit, in spite of the recent advances in correcting aberrations. There are several reasons limiting the resolution of electron microscopes, such as spherical aberrations in TEM. However, the most fundamental reason preventing electron microscopy from reaching the fundamental resolution limit is the interaction between electrons[16,17], which is called *the space-charge* effect. This effect is especially important in microscopes of lower electron velocities or larger currents, both happening in SEMs. With the gradual improvements in electron microscopes, and especially with better techniques for aberration corrections, this fundamental resolution limit (arising from the space charge effect) is quickly becoming the dominant barrier for further improvements[16–21]. Of course, when the density of the electrons in the beam is low enough, this effect becomes negligible. However, working with one electron at a time[17] implies longer integration times in the detection process to obtain a reasonable SNR. This space charge effect is of even greater importance in low voltage electron microscopes, which are becoming more popular recently. There, electron-electron interaction is already preventing an even lower acceleration voltages[22].

In principle, shaping the quantum wavefunction of electrons has the potential to improve the performance of traditional electron microscopes. However, thus far most electron microscopes have been using only low electron currents, working with one electron at a time, where the space-charge effect is negligible. The intrinsic reason for that is that electron-electron interactions cause repulsion between the electrons, which broadens the EBeam, which in turn considerably hampers the resolution. For this reason, most electron microscopes rely on relatively low currents, which can be fully described by single electrons.

Here, we develop a novel quantum technique to compensate for the repulsion between electrons and generate a high density EBeam while maintaining the resolution of a single electron. We propose to do that by proper shaping of the quantum wavefunction of multiple electrons, so as to counteract both the repulsion and the diffraction-broadening. To find such a non-diffracting multi-electron beam, we formulate the multiple electron Schrödinger equation, which is nonlinear due to the interaction among electrons. Then, we solve for the wavefunction that preserve its shape in time. Our technique facilitates the use of EBeams made up of multiple electrons without compromising on the spatial resolution. It enables higher SNR with short integration time, by working with high density EBeams while exhibiting spatial resolution equal to the resolution of a single-electron. As such, it can contribute to all electron beams applications and experiments, such as electron microscopy, free electron lasers, electron beam lithography, accelerators, etc. Additionally, future studies can lead to fabrication of masks for heavier charged particles, such as beams of ions, protons and even muons. Our scheme still applies in such cases and gives significantly better results because heavier charged particles have much lower velocity for the same acceleration field.

The multiple-electron Schrödinger equation contains terms of interaction between electrons. Those terms change the Schrödinger equation to become a nonlinearly-coupled set of wave equations. The full many-body problem is complex because the number of degrees of freedom is proportional to the number of electrons, which is computationally intractable (for a large number of electrons) with classical computers. Thus, we apply a mean-field approach (the Hartree approximation) to reduce the problem to a smaller number of nonlinearly-coupled wave equations, which can still describe a large number of electrons. Mean-field approximations are commonly used for free-electrons, such as plasma[23,24], EBbeams[25] and many other fermionic systems[26]. From these coupled equations, we find the nonlinear non-diffracting EBeam wavefunctions that can also carry OAM. In other words, we find multi-electron vortex beams that preserve their shape. Generally, these solutions are not square-integrable, similar to the Bessel and Airy beams, hence generating them in a physical setting implies truncating their wavefunctions, which implies that they remain non-diffracting only for a finite range. However, simulating their evolution (propagation dynamics) shows that this non-diffraction range can be very large, and proves their robustness to noise and to deviations from non-ideal launch conditions. Finally, comparing the non-diffraction range of our shape-preserving wavefunctions with that of a Gaussian EBeam (which is roughly the wavefunction naturally occurring in electron microscopes) and with a Bessel EBeam (the non-diffracting analogue for a single electron), truncated by the same aperture, we observe substantially larger range of our multi-electron wavefunctions. This paves the way for using properly shaped multi-electron beams in electron microscopy as well as in a variety of other applications.

Thus far, wavefunction shaping of EBeams[7,8,10–12] have only considered EBeams comprising of a single electron. As such, these methodologies are inapplicable for high density electrons beams, where electronic repulsion is appreciable. At the same time, high density EBeams are encountered in a variety of applications, ranging from accelerators, high current and low voltage electron microscopy to high intensity X-ray sources (e.g., FEL) and much more. In a similar vein, controlling and shaping high density EBeams is also important from the basic science view point.

The repulsion between electrons renders the beam diffraction to be density dependent, thereby making the problem nonlinear. That is, while a sufficiently dilute EBeam is described by the linear Schrödinger equation, the physics becomes considerably more complex when many-body interactions take place. This draws a fundamental difference between electron beams and electromagnetic beams: at intensities lower than $10^{22}$ Watts/cm$^2$, where vacuum QED effects are negligible[27], photons do not directly "interact" with one another, whereas electrons inevitably always interact with one another.

Importantly, in addition to Coulomb repulsion, EBeams are also subject to spin-spin interaction, and in case of beams carrying OAM, also to spin-orbit interaction, thereby adding additional complexity to the dynamics[28]. However, under the typical parameters of electron microscopes (primarily that all features are much larger than the Compton wavelength), the spin-spin and spin-orbit interactions are negligible compared to the electrostatic potential energy (see Supplementary Information for details).

The exact Schrödinger equation contains a nonlinear set of coupled equations whose number is of the number of electrons. To make the problem tractable, we approximate the full multi-electrons Hamiltonian by the Hartree Hamiltonian, which is an effective mean-field Hamiltonian. This approach assumes that the influence of the fermionic nature of the electrons (the exclusion principle) is very weak[29]. Hence, we are allowed to take the simplest case where all the electrons have the same wavefunction, as happens naturally in electron microscopes. We also restrict the wavefunction to be cylindrically symmetric while allowing it to carry OAM. This wavefunction is therefore of the form

$$\psi(\mathbf{r},t) = \frac{1}{a_0}\phi(\rho)e^{il\theta}\frac{e^{ikz-i\omega t}}{\sqrt{L}}. \qquad (1)$$

Where, $a_0$ is Bohr's radius, $l$ is the OAM, and $k$ is the wavenumber in the $z$ direction. The normalization factor, $\sqrt{L}$, sets the characteristic length scale ($z$) within which the wavefunction is significant (the so-called uncertainty length). Although this factor cancels out later on, this length scale is useful for estimating the strength of the effects involved (see Supplementary Information). The time evolution of the wavefunction $\psi$, according to the Hartree Hamiltonian[30], is

$$-i\hbar\partial_t\psi(\boldsymbol{r},t) = -\frac{\hbar^2}{2m}\nabla^2\psi(\boldsymbol{r},t) + \frac{Ne^2}{4\pi\varepsilon_0}\left(\int\frac{|\psi(\boldsymbol{r}',t)|^2}{|\boldsymbol{r}-\boldsymbol{r}'|}d^3\boldsymbol{r}'\right)\psi(\boldsymbol{r},t). \quad (2)$$

Where $\hbar$ is the reduced Planck constant, $m$ and $e$ are the mass and charge of the electron respectively, $\varepsilon_0$ is the vacuum permeability and $N$ is the total number of electrons in the EBeam. This equation is known as "Choquard equation"[31]. The second term in the right hand side of Eq. 2 resembles an effective potential. Hence, we define

$$U(\boldsymbol{r},t) = 2Na_0\int\frac{|\psi(\boldsymbol{r}',t)|^2}{|\boldsymbol{r}-\boldsymbol{r}'|}d^3\boldsymbol{r}'. \quad (3)$$

Substituting Eq. (3) in Eq. (2) we get the following coupled equations,

$$-i\hbar\partial_t\psi(\boldsymbol{r},t) = \frac{\hbar^2}{2m}\left(-\nabla^2 + \frac{1}{a_0^2}U(\boldsymbol{r},t)\right)\psi(\boldsymbol{r},t) \quad (4.1)$$

$$\nabla^2 U(\boldsymbol{r},t) = -8\pi Na_0|\psi(\boldsymbol{r},t)|^2. \quad (4.2)$$

Substituting the wavefunction from Eq. 1 as a source term in Eqs. 4, we get that the effective potential also has rotational symmetry, and as such it depends only on $\rho$, such that $U(\boldsymbol{r},t) = U(\rho)$. We now look for a solution that is shape-preserving, namely, we seek a solution whose expectation value does not vary in time. This allows the separation into two coupled nonlinear differential equations,

$$-\left(\frac{1}{\rho}\partial_\rho(\rho\partial_\rho) - \frac{l^2}{\rho^2}\right)\phi(\rho) + \frac{1}{a_0^2}U(\rho)\phi(\rho) = 0 \quad (5.1)$$

$$\frac{1}{\rho}\partial_\rho(\rho\partial_\rho)U(\rho) = -\frac{8\pi n}{a_0}|\phi(\rho)|^2. \quad (5.2)$$

where, $n$ is the density of electrons per unit distance. These equations resemble the Newton–Schrödinger model[32,33], which is often used in General Relativity to describe the dynamics of wavefunctions under the gravitation potential they themselves induce. This set of equations also resemble the equations used to describe the dynamics of optical beams in the presence of the highly nonlocal optical thermal nonlinearity, which supports solitons[34] and their long-range interactions[35]. Such an optical system was recently used to emulate effects predicted in General Relativity, and discover new phenomena[36]. Interestingly, this system also resembles the model describing long-range interactions between cold atomic dipoles[37], which also give rise to solitons and related phenomena. These "nonlocal solitons"[34,35] and their counterparts in cold dipoles[37] resemble the shape-preserving multi-electron wavepackets found here, as solutions to Eqs. 5. However, whereas in the Newton–Schrödinger model the force is always attractive, the force here is always repulsive. This means that, while the nonlinear non-diffracting wavepackets in the Newton–Schrödinger model are solitons, and are therefore localized and square-integrable[34], we expect the localized non-diffracting solutions of Eqs. 5 to be not square-integrable. Intuitively, seeking localized solutions for Eqs. 5 resembles searching for non-diffracting beams in self-defocusing thermal optical nonlinearities, which fundamentally cannot support bright solitons but can support dark solitons[38] and also localized non-diffracting wavepackets that are not square integrable (e.g., nonlinear Bessel-like beams[39]).

To solve these differential equations and find a shape-invariant solution, we need to determine the initial conditions. The wavefunctions we seek have rotational symmetry with respect to the propagation axis ($z$), hence so does the effective potential, which yields $U'(\rho = 0) = 0$. Substituting this condition into Eq. (5.1) at the vicinity of $\rho = 0$ gives the Bessel equation, whose solution is $\phi(\rho) \sim \alpha J_l(k_T \rho)$ (the other solution, $Y_l(k_T \rho)$, is unphysical because it diverges at $\rho = 0$). Using $\phi(\rho)$ in Eq. (5.1) leads to $U(\rho = 0) = -k_T^2$, where $k_T$ is a real positive number which corresponds to the transverse momentum. Another initial condition is provided by the normalization requirement $2\pi \int |\phi(\rho)|^2 \rho d\rho = 1$, where the integral boundaries correspond to the beam aperture. Hence, for a given aperture we get a continuous set of solutions, determined by the free parameter $k_T$, which can vary between 0 and infinity.

Figure 1a shows an example of the radial wavefunction $\phi(\rho)$ of a shape-invariant solution with zero OAM. It is instructive to compare this wavefunction (which is a shape-invariant solution of the nonlinear equation) to the Bessel function, which is a shape-invariant solution of the linear equation describing the evolution of a single electron[12]. Figure 1a shows this comparison, with the same value of $k_T$. In the vicinity of $\rho = 0$, the shape-invariant solution coincides with the Bessel function $J_0(k_T \rho)$, as discussed in the previous paragraph, but for large $\rho$ it displays denser lobes, which means that the wavefunction carries higher transverse momentum. The reason for this is that the effective potential representing the repulsion between electrons, $U(\rho) = -k_T^2(\rho)$, is increasing with $\rho$, which implies that higher transverse momentum (denser oscillations) is required to compensate for the repulsion at higher $\rho$ values. An interesting case is shown in Fig. 1b, which presents the radial function found for $k_T = 0$ (with $\phi'(0) = 0$ and $\phi''(0) = 0$), which yields the upper limit to the width of the main lobe of the radial wavefunction of the shape-invariant solutions of Eqs. 5. This solution does not have a corresponding linear solution, because the Bessel function becomes a constant (corresponding to a plane wave) for $k_T = 0$. Figure 1c displays several radial wavefunctions of shape-invariant solutions that carry OAM (with the same value of $k_T$ and aperture size as in Fig. 1a). The blue, green, red and cyan curves correspond to OAM of zero, one, three and five, respectively.

The non-diffracting wavefunctions (the solutions of Eqs. 5) can be generated by passing the EBeam through a binary holographic mask as in[6,7,10–12], or through a phase mask imprinting the actual phase distribution of the shape-invariant wavefunction[8], which shapes the electron wavepacket directly. The holographic mask has the following transmission function

$$T_{holographic\ mask} = \left|\mathcal{F}\{\phi(\rho)e^{il\theta}\} + e^{ik_h\rho\cos\theta}\right|^2 \qquad (6.1)$$

Where $\phi(\rho)e^{il\theta}$ is the non-diffracting wavefunction (solution of Eqs. 5), F is its Fourier transform, and $e^{ik_h\rho\cos\theta}$ is a plane wave acting as a reference for the hologram. This transmission function is attains a binary shape, as in[6,7,10–12] according to

$$T^{binary}_{holographic\ mask} = \begin{cases} 1, & T_{holography} > threshold\ and\ \rho < \rho_{max} \\ 0, & else \end{cases}. \qquad (6.2)$$

As an example, consider the mask shown in Fig. 2, which is the binary holographic mask required for shaping the wavefunction shown in Fig. 1a. Figure 2b shows the far-field diffraction pattern obtained when passing a plane wave through this mask. The central waveform is the zeroth-order diffraction pattern. The non-diffracting wavefunction is on the left, and its complex conjugate appears on the right, corresponding to +1 and -1 diffraction order. Figure 2c singles out the first diffraction order, showing a very good agreement with Fig. 2d, which represents the absolute value of $\phi(\rho)$ – the non-diffracting wavefunction defined by Eqs. 5.

At this point it is important to simulate the propagation dynamics of the multi-electron beams we have found, which are meant to be shape-invariant for some finite propagation range, and compare them to the evolution of the Bessel beam (the diffractionless solution for a single-electron) and to the evolution of Gaussian EBeam. Figure 3 presents the simulated propagation dynamics of several wavefunctions, displaying the density of the EBeams as a function of $\rho$ and the propagation distance z. In all of these examples, the acceleration voltage is $20kV$ (typical SEM energies) and the current is $50\mu A$ - which is considered a very high current in microscopes, meant to highlight our findings (Ref [40] presents the use of such high current together with coherent tip and coherent EBeam).

The simulation method used for 2D+1 quantum system (described in Eqs. 4) is the Beam Propagation Method (BPM) (or Split Step Fourier method), with an addition of a procedure calculating the potential $U$ at each step. Specifically, given an initial wavefunction of some shape $\psi(x, y, t)$, we calculate the potential $U(x, y, t)$ by solving numerically Eq. 4.2, for the initial conditions described in the Supplementary Information. Then, we use this potential in Eq. 4.1 to calculate the beam in the next step $(x, y, t + dt)$, and so on.

In coming to examine the diffraction–broadening effects during propagation of these beams, we recall that the resolution of EBeam microscopes is determined by the region of high density of electrons (high current density), which is crucial in electron microscopy. It is therefore natural to examine the width of the region of high electron density in the beam. However, observing Fig. 3, we notice that our non-diffracting multi-electron wavefunctions exhibit diffraction effects that are fundamentally different than diffraction of Gaussian beams. Namely, whereas in Gaussian beams the width expands monotonically with distance, for the non-diffracting wavefunctions – when they are launched through an aperture sufficiently far away from their main lobe - the of the main lobe maintains its width and shape virtually indefinitely in spite of the truncation. Instead, for the non-diffracting wavefunctions the zeros around the main lobe fill up, and the contrast between the main lobe and the secondary lobes vanishes, thereby reducing the resolution, while the full-width-half-maximum of the main lobe of the shape-preserving multi-electron beam (Fig. 3e) varies very little (unlike the truncated Bessel beam whose FWHM varies considerably, as shown in Fig. 3d). Actually, the resolution of an EBeam is determined by the size of the region of high intensity (region of high probability). We therefore define a measure for effective width of the main lobe, as the second moment of the electron density, measured in the main lobe region (defined by the zero around it), as

$$w = \frac{\sqrt{\iint_{Main\ lobe} |\psi(x,y)|^2 (x^2+y^2)\, dxdy}}{\sqrt{\iint_{Main\ lobe} |\psi(x,y)|^2\, dxdy}}. \qquad (7)$$

For the Gaussian wavefunction (which has no zeros), this effective width is simply the second moment. In all the examples in Fig. 3, the effective width is $w =$

$16 nm$. In this vein, we define the range of non-diffraction $L_d$ as the distance for which the effective width is increased by a factor of $\sqrt{2}$. The non-diffraction ranges of the various beams in Fig. 3 are marked by the horizontal dashed line in each panel.

Figure 3 shows the simulated evolution of several wavefunctions launched as initial conditions for solving Eqs. 5, where the single-electron cases (Figs. 3a,b) does not include the effective potential term ($U$), whereas the multi-electron cases (Figs. 3c,d,e,f) include the nonlinear term representing the repulsion among electrons. Figures 3a and 3b present the propagation of single-electron beam of initial Gaussian and Bessel wavefunctions, respectively. Both wavefunctions launched through an aperture of $140 nm$, which has no effect on the Gaussian beam but truncates the Bessel beam after 10 oscillatory lobes. The single-electron Gaussian beam exhibits fast diffraction ($L_d = 4.2 \mu m$), while the single-electron Bessel beam preserves its shape for a very large range distance ($L_d = 141.6 \mu m$). Figures 3c,d present the corresponding cases for EBeams comprising of multiple electrons. Clearly, the Gaussian multi-electron beam (Fig. 3c) expands faster ($L_d = 3 \mu m$) than the single-electron beam (Fig. 3a), due to the repulsion among electrons. Likewise, the multi-electron Bessel beam (Fig. 3d) also expands faster ($21 \mu m$) compared to the single-electron case which ideally (had it not been truncated) would remain non-diffracting indefinitely (Fig. 3b). Clearly, the Bessel function is not a suitable non-diffracting solution of the nonlinear evolution equation (Eq. 2). On this background, Fig. 3e presents the propagation dynamics of our nonlinear shape-preserving wavefunction shown in Fig. 1a. Had this wavefunction not been truncated by the aperture, it would have preserved its shape indefinitely, in spite of the repulsion among electrons. The aperture truncates the wavefunction after 17 oscillatory lobes, and consequently causes the diffraction effects shown in Fig. 3e. Figure 3f shows the evolution of the same wavefunction in the presence of additive Gaussian noise. To highlight the robustness of our findings, Fig. 3f simulates the extreme case where the total noise current is equal to the total current carried by the aperture beam of Fig. 3e, and is uniformly distributed in real-space. As shown there, the noise does not have any noticeable effect on the evolution of the shape-preserving wavefunction of the multi-electron beam, and that the propagation dynamics is robust to deviations from non-ideal launch conditions.

Examining the diffraction of the multi-electron beam of Fig. 3e preserves it exact shape up to a propagation distance of $L_d = 96 \mu m$, and then the main lobe and the entire structure fade away quickly, within a short distance. Clearly, the non-diffraction range of the multi-electron shape-preserving beam of Figs. 1a and 3e is 5 times larger than the non-diffraction range of the multi-electron Bessel beam of Fig. 3d. Actually, the non-diffraction range of the multi-electron shape-preserving beam of Fig. 3e is closer to the corresponding range of the single-electron Bessel beam of Fig. 3b.

Altogether, as highlighted by Fig. 3, the wavefunction we find by seeking propagation-invariant solutions to the multiple electron Schrödinger equation is indeed shape-preserving. It overcomes the repulsion among electrons and the natural tendency of diffraction broadening inherent in the Schrödinger equation. Moreover, when this wavefunction is launched from a finite aperture, it preserves its shape for a distance close to the range of the corresponding single-electron Bessel beam launched from the same aperture, in spite of the fact that the multi-electron beam carries very high current – corresponding to 74,000 electrons per cm. This fact implies that shape-preserving multi-electron beams can be launched from the very same apertures and under the same noise conditions as Bessel EBeams are launched today in electron microscopes and in other applications.

Figure 4 presents a quantitative comparison in the performance between our shape-preserving multi-electron wavefunction and multi-electron Bessel and Gaussian beams, all carrying zero OAM. Figure 4 shows the non-diffraction range as a function of the effective width of the initial beam, with acceleration voltage of $20kV$ and current of $50\mu A$. The shape-preserving multi-electron beam (solid blue curve) performs remarkably better than the Gaussian beam (dotted blue curve) and also considerably better than the Bessel beam (dashed blue curve). The red curves display the current carried by the main lobe, solid for our shape-preserving wavefunction and dashed for the multi-electron Bessel beam. As shown there, the main lobe of the non-diffracting wavefunction carries more current than the Bessel beam, in addition to its better performance in terms of the non-diffraction range.

Interestingly, the dashed blue curve in Fig. 4 (describing the non-diffraction range of the Bessel beam) has a turn at $4.2nm$. We refer to this turn as the critical width (black dot-dashed vertical line), below which the performance of our shape-preserving wavefunction coincides with that of multi-electron Bessel beam (launched from the same aperture). We find that this critical width decreases as we increase the electron density in the beam (increasing the current), and it can go below $1nm$. This is because, in the region of very narrow multi-electron beams, that corresponds to $k_T$ much larger than the inverse of the critical width, the potential satisfying Eq. 5.2 with initial condition $U(\rho = 0) = -k_T^2$ goes to a constant. This makes the wavefuction obeying Eq. 5.1 coincide with the Bessel function, hence their performances (non-diffraction range and current carried by the main lobe) coincide as well. In the other regime, for main lobe widths larger than the critical width, our shape-preserving wavefunction performs much better than the Bessel beam, as highlighted by Fig. 4.

Another important feature that can be seen in Fig. 4 is that the width of the nonlinear shape-preserving wavefunction is bounded from above, at the blue dot (henceforth referred to as the maximal width), where the main lobe is over-wide. This over-wide wavefunction also marks the upper limit on the current. The reason for the existence of this upper limit is that the interference effects caused by the shape of our multi-electron wavepacket can balance the beam's self-repulsion and diffraction only up to a certain electron density, above which the repulsion is too strong to be compensated by the predesigned interference effects. This upper limit point occurs for $k_T = 0$, which corresponds to the beam with the widest main lobe (Fig. 1b). The main lobe of this maximum-width non-diffracting wavefunction carries considerably higher current than the main lobe of the corresponding multi-electron Bessel beam.

Recalling that the ideal (non-truncated) multi-electron beams are not square-integrable (i.e., they carry infinite power), like the Bessel beam, their diffraction properties are also determined by the aperture at the launch plane. The role played by the aperture on defining the non-diffraction range $L_d$ is discussed in the Supplementary Information and the figure therein.

Finally, it is interesting to study the propagation evolution of the shape-invariant multi-electrons beams that do carry OAM, such as those shown in Fig. 1c. Figures 5a-d present the simulated propagation of such wavefunctions (while neglecting the spin-spin and spin-orbit interaction; see Supplementary Information), with acceleration voltage of $20 kV$ and current of $50 \mu A$. Figures 5a,b show the propagation of initial wavefunction of Gaussian and Bessel shapes, with OAM=1, respectively. Figure 5c presents the propagation of our shape-preserving wavefunction with OAM=1. Similar to the case without OAM, the non-diffraction range of the shape-preserving beam (Fig. 5c) is much larger than the non-diffraction range of the Bessel beam (Fig. 5b). The performance of these multi-electron beams carrying OAM is similar to the trend shown in Fig. 4: the non-diffraction range is order of magnitude larger than for multi-electron Gaussian and Bessel beams, and is in fact similar to the non-diffraction range of the respective single-electron Bessel beam launched from the same aperture.

Before closing, it is important to discuss the potential applications of our findings. Clearly, the non-diffracting multi-electron beams found here have inherent fundamental importance – similar to the impact made by the optical Bessel beam (which was the first non-diffracting beam discovered). In addition, the concept of non-diffracting multi-electron beams also has profound potential for applications, especially in electron microscopy. Specifically, the current in electron microscopes is proportional to the density of electrons. Converting to the spatial density only requires dividing by the velocity; hence, the nonlinear term in Eqs. 5 is proportional to the current divided by the square root of the acceleration voltage (for nonrelativistic EBeams). Therefore, significant repulsion among electrons can arise either from high current EBeam or from low acceleration voltage. SEM and STEM work by focusing the EBeam on the sample under study[42], hence the resolution in both of them is determined by the diameter of focused spot. Naturally, employing SEM and STEM in the high current regime (tens to hundreds of microAmperes) would cause loss of resolution due to the repulsion[43], which is exactly what our technique can counteract. Under realistic parameters of current SEM technology, our technique can increase the current density by at least factor $10^6$ while maintaining resolution of 1nm (assuming that the high current does not damage the SEM components and the sample). In an alternative application, our method can be exploited in low-voltage electron microscopy (LEEM)[44,45] to counteract the repulsive loss of resolution (that is especially significant due to the very low velocities). Likewise, our technique can be very important to microscopes working with ultrashort pulses of electrons (ultrafast electron microscopy)[19,46], where the electron density could be very high due to the ultrashort duration of the pulse. For this application, it should be noted that applying our approach to ultrashort pulses would require dealing with a broad spectrum of electron energies and not just a mono-energetic EBeams (as we analyzed here). We leave this to future research, although it is clear that our approach can be modified to shape both the spatial and the frequency spectra associated with the broad spectrum of ultrashort pulse EBeams. Additionally, our technique can also be used to increase the yield of the spatially-coherent electrons passing through the condensers in electron microscopes. Here, normally only a small part of the electron wavefunction can pass through the condenser's aperture (acting as a spatial filter), in part due to the space charge effect; hence, our method can lead to increasing the flux of electrons that are simultaneously localized in real space and in momentum space. Physically, this can be achieved by positioning the mask immediately after the condenser lens, or possibly even add a quadratic phase to the mask so as to make it serve as a both the lens and the generator of the non-diffracting wavefunction.

To conclude, we have shown that the wavefunctions of multi-electron beams, or any other beams of charged particles, e.g. protons, muons and ion beams, can be properly designed to compensate for both space-charge (self-repulsion) effects and diffraction broadening, and can even carry orbital angular momentum. Our simulations predict that our shaped non-diffracting beams perform remarkably better than the multi-electron Bessel and Gaussian EBeams. The design methodology presented here finds applications in electron microscopy, electron beam lithography, accelerators and a variety of other applications. Using our shaped multi-electron beams in low energy and high current microscopes, one can still achieve high resolution despite the repulsion among the electrons. Essentially, what we suggest here can resolve the space-charge field effects that appear in all technologies using beams of multiple electrons. However, further research is necessary to understand the degree of coherent possible in high current electron beams. Finally, we recall the resemblance of our model for multi-electron beams to the Newton-Schrödinger model known from General Relativity (with the exception that the force in our EBeam is repulsive, whereas the force in the Newton-Schrödinger model is attractive). We also note the similarity of our non-diffracting multi-electron beams to solitons in nonlocal nonlinear media in optics and in cold atomic dipoles. These resemblances raise a series of intriguing questions, among them: the existence of dark solitons made of multi-electron beams, and long-range interactions among such self-trapped entities.

**Figure 1**

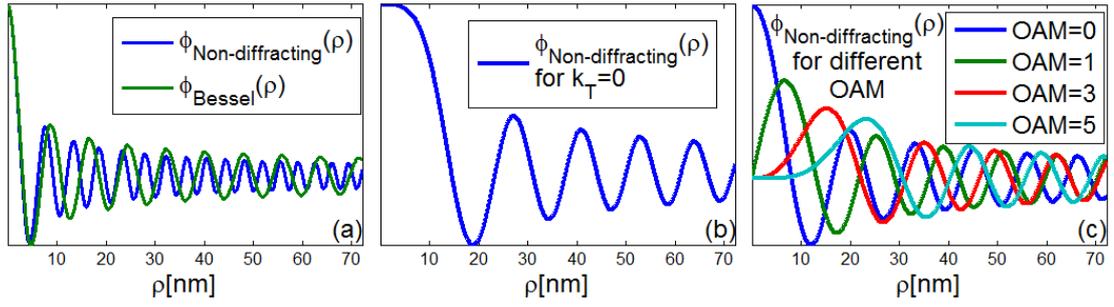

**Fig. 1**: **Radial part of the shape-invariant wavefunctions of multi-electron beams.** (a) Radial wavefunction of the beam with zero OAM (blue), compared with the Bessel function which is the corresponding wavefunction of a single-electron beam. (b) The unique over-wide radial wavefunction obtained for $k_T = 0$, which does not allow OAM, and does not have a corresponding linear solution. (c) Radial wavefunctions of multi-electron beams carrying OAM = 0,1,3,5 (blue, green, red and cyan, respectively).

**Figure 2**

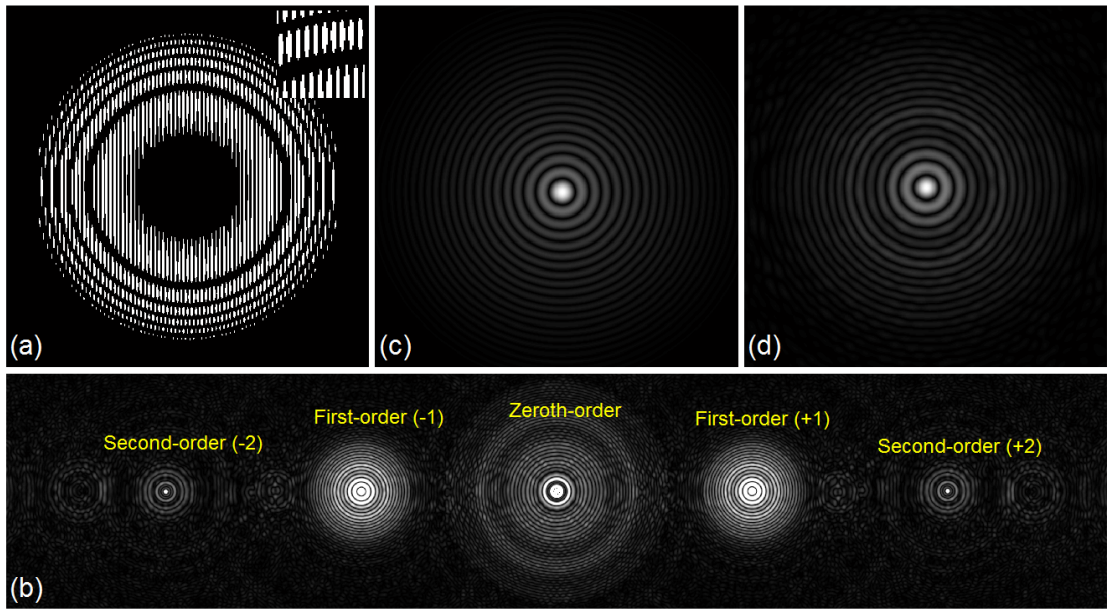

**Fig. 2**: **Binary Holographic mask that generates the shape-invariant wavefunction of multi-electron beams and the diffraction pattern it generates.** (a) Transmission function of the binary mask. (b) The far-field diffraction pattern generated by passing a plane wave through this mask. The center pattern is the zeroth-order diffraction pattern. The pattern on the left (right) of the zeroth order corresponds to the +1 (-1) diffraction order. This far-field pattern shows that the diffraction patterns can be cleanly separated from one another, and that the ±1 orders can be used to generate the non-diffracting wavefunction. (c) and (d) Comparison between the diffraction pattern obtained from the +1 order and the desired wavefunction for which the mask is designed. This demonstrates that a very good approximation of the desired wavefunction can be generated even with a binary mask fabricated with present technologies (of course adding a phase mask[8] would give an even better result).

**Figure 3**

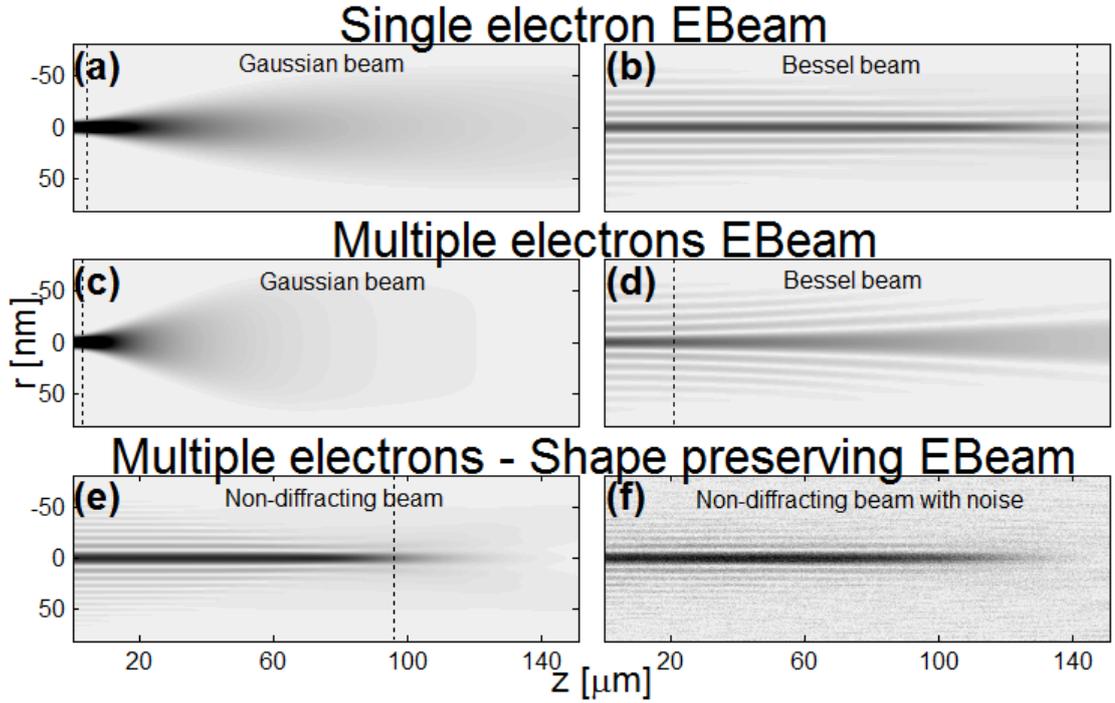

**Fig. 3: The density of various single-electron and multi-electron beams as a function of propagation distance.** (a,b) Propagation of a single-electron beam (which does not have self-repulsion) with initial Gaussian and Bessel wavefunctions, respectively. (c,d) Propagation of the respective cases for EBeams containing multiple electrons. The repulsion among the electrons makes the Gaussian EBeam diffracts faster than in (a), while the Bessel broadens considerably and is no longer shape preserving. (e) Propagation of the non-diffracting multi-electron beams with the wavefunction of Fig. 1a. The beam preserves its shape for a large distance in spite of the repulsion among electrons. (f) Same as (e) with added noise, exhibiting robustness to high level of noise (uniformly-distributed Gaussian noise that carries the same power as the beam). The dashed lines in (a)-(e), indicate the range of non-diffraction of each beam. Here, the EBeams are accelerated by voltage of $20 kV$, having beam current of $I = 50 \mu A$ (for the multi-electron beams), and effective width of $16 nm$ (Eq. 7).

**Figure 4**

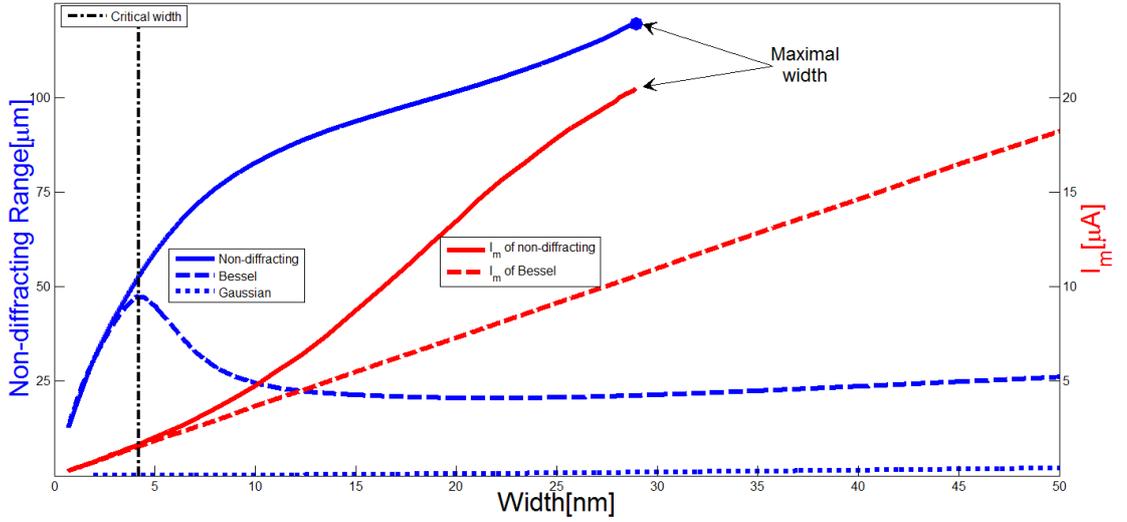

**Fig. 4: Non-diffraction range (blue curves) and effective current (red curves) vs. beam width, for the shape-preserving wavefunctions (solid curves), and for multi-electron Bessel (dashed curves) and Gaussian (dotted curve) beams with zero OAM.** The effective width of the shape-preserving multi-electron beam is bounded from above at the blue dot, where the main lobe is over-wide, carrying the upper limit on the current. This upper limit occurs because the interference of our shaped wavepacket can balance the beam self-repulsion and diffraction only up to a certain value, above which the beam spread is too strong for the predesigned interference effects (arising from the structure of the beam) to compensate for it. For beams of effective width much narrower than the critical width (marked by the dashed vertical line at $4.2 nm$) the shape-preserving multi-electron wavefunction coincides with the Bessel function. This critical width decreases as the current is increased and can go below $1 nm$. Here, the EBeams are accelerated by voltage of $20 kV$, having beam current of $I = 50 \mu A$.

**Figure 5**

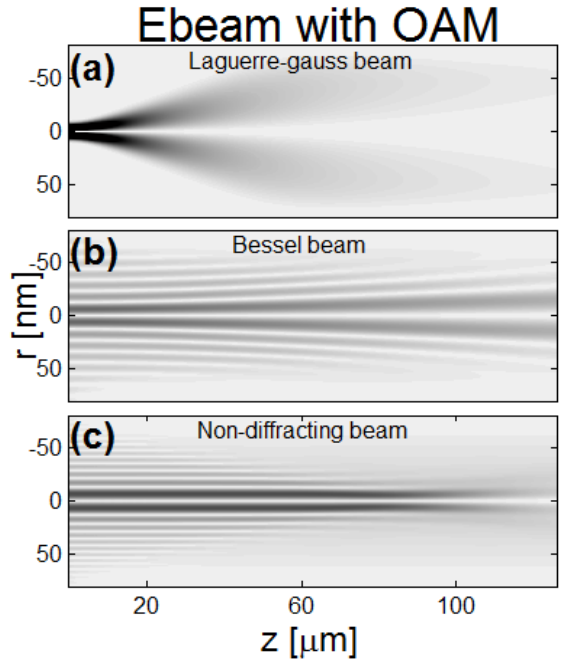

**Fig. 5: Evolution of multi-electron beams with nonzero OAM, as a function of propagation distance.** (a,b) Propagation of a multi-electron beam with OAM=1 and initial Laguerre-gauss and Bessel wavefunctions, respectively. The repulsion among the electrons makes the Laguerre-gauss EBeam diffracts very fast, while even the Bessel beam broadens considerably. (c) Propagation of the non-diffracting multi-electron beams with OAM=1 and the wavefunction represented by the green curve in Fig. 1c. The beam preserves its shape for a large distance in spite of the repulsion among electrons.

# Non-Diffracting Electron Vortex Beams Balancing Their Electron-Electron Interactions

## Supplementary Information

Maor Mutzafi[1], Ido Kaminer[2,1], Gal Harari[1] and Mordechai Segev[1]

1. Physics Department and Solid State Institute, Technion, Haifa 32000, Israel
2. Department of Physics, Massachusetts Institute of Technology, 77 Massachusetts Avenue, Cambridge, Massachusetts, USA

## 1. Derivation of the equation of the non-diffracting beam

This section described the derivation of the equation governing the propagation of the multi-electron beam. The Schrödinger Equation for a system of multiple electrons under the Hartree (mean-field) Hamiltonian, is as follows:

$$-i\hbar\partial_t \psi_i(\boldsymbol{r},t) = -\frac{\hbar^2}{2m}\nabla^2\psi_i(\boldsymbol{r},t) + \frac{e^2}{4\pi\varepsilon_0}\sum_{j=1,j\neq i}^{N}\left(\int \frac{|\psi_j(\boldsymbol{r}',t)|^2}{|\boldsymbol{r}-\boldsymbol{r}'|}d^3\boldsymbol{r}'\right)\psi_i(\boldsymbol{r},t). \quad \text{(viii)}$$

Where, $\psi_i$ and is the wavefunction of the $i^{\text{th}}$ electron, $\hbar$ is the reduced Planck constant, $m$ and $e$ are the mass and charge of the electron respectively, $\varepsilon_0$ is the vacuum permeability and $N$ is the total number of electrons in the EBeam.

Next, we assume that all the electrons have the same wave function, as reflected by Eq. 1 in the paper. We also assume that the "self-interaction" (when $j=i$) is negligible, which is the case when the beam is very dense ($N$ is a large number). Proceeding to substitute the wavefunction from Eq. 1 and the potential form from the main text, we recover Eqs. 5 therein, which we recall here:

$$-\left(\frac{1}{\rho}\partial_\rho(\rho\partial_\rho) - \frac{l^2}{\rho^2}\right)\phi(\rho) + \frac{1}{a_0^2}U(\rho)\phi(\rho) = \left(\frac{2m}{\hbar^2}E - k^2\right)\phi(\rho) \quad \text{(ix)}$$

$$\frac{1}{\rho}\partial_\rho(\rho\partial_\rho)U(\rho) = -\frac{8\pi n}{a_0}|\phi(\rho)|^2 \quad \text{(x)}$$

It is now convenient to define $U(0) = a_0^2 \left(\frac{2m}{\hbar^2} E - k^2\right)$, which simplifies Eq. (ii) to Eq. 4.1 from the paper. The normalization requirement is:

$$2\pi \int |\phi(\rho)|^2 \rho d\rho = 1 \qquad \text{(xi)}$$

The initial conditions for the nonlinear set of equations are:

$$\begin{cases} \phi(0) = \alpha \\ \phi'(\varepsilon) = \alpha k_T J_l'(k_T \varepsilon) \\ U(0) = -k_T^2 \\ U'(0) = 0 \end{cases} \qquad \text{(xii)}$$

where, $\alpha$ is determined from the normalization requirement. The fact that $|\phi(\rho)|^2$ is symmetric in space implies that $U'(0) = 0$, hence the only free parameter in Eqs. (ii) is $k_T$ - which can vary between 0 and infinity. At the vicinity of $\rho = 0$, Eq. (ii) gives the Bessel equation, whose solution is $\phi(\rho) = \alpha_1 J_l(k_T \rho) + \alpha_2 Y_l(k_T \rho)$. However, $Y_l(k_T \rho)$, is unphysical because it diverges at $\rho = 0$. As such, we are left with the first term only, which is the reason why $k_T$ is the only remaining degree of freedom.

The electron density on the $z$ axis of an EBeam (as appears in Eq. 4.2 in the paper) can be derived from the current $I$, and the acceleration voltage $V$:

$$n = \frac{dN}{dz} = \frac{1}{e}\frac{dq}{dz} = \frac{1}{e}\frac{dq}{dt}\frac{dt}{dz} = \frac{I}{ev} = \frac{I}{\sqrt{V}}\sqrt{\frac{m}{2e^3}}. \qquad \text{(xiii)}$$

As a side note, we would like to add that, while the above treatment is non-relativistic, it can be directly extended to a fully relativistic quantum formalism. In case the propagation is limited to small angles (paraxial EBeams), the Schrödinger Equation only needs to be changed by multiplying the mass by the relativistic gamma, and decreasing the interaction terms by the same factor. In any case, the non-relativistic equation above is a very good approximation for the parameters we simulate in the paper.

## 2. Neglecting the spin-spin and spin-orbit interaction

This approach is along the lines of previous work addressing related questions (see, reference [24] in the paper) that showed that, in most standard EBeam conditions, the coulomb interaction dominates over any spin-related effect. Here, the energy of the spin-spin and spin-orbit interaction is as follows:

$$E_{spin} \sim \frac{Ne^2 h^2}{4\pi\varepsilon_0 m^2 c^2} \left\langle \frac{1}{r^3} \right\rangle \qquad (xiv)$$

While the energy related to the potential energy of Coulomb interaction is:

$$E_{e-e} \sim \frac{Ne^2}{4\pi\varepsilon_0} \left\langle \frac{1}{r} \right\rangle \qquad (xv)$$

The ratio between them is:

$$\frac{E_{e-e}}{E_{spin}} \sim \frac{m^2 c^2}{h^2} \frac{\langle \frac{1}{r} \rangle}{\langle \frac{1}{r^3} \rangle} = \frac{L_{typical}^2}{\lambda_c^2} \qquad (xvi)$$

Where, $L_{typical}$ is a typical length scale in the system and $\lambda_c$ is the Compton wavelength. The EBeams we consider have two typical lengths: the aperture size in the $x, y$ plane, and the average distance between electrons ($1/n$) along the $z$ axis. Both are typically $1nm - 1\mu m$ or even larger, while the Compton wavelength is $\lambda_c = 2.4pm$. This means that the spin-spin and spin-orbit interactions are negligible in this system.

$$\frac{E_{e-e}}{E_{spin}} \sim \frac{L_{typical}^2}{\lambda_c^2} \gg 1. \qquad (xvii)$$

In a similar vein, previous work that compared the spin-spin interaction and the coulomb repulsion led to similar conclusions (ref. [24] in the paper): in most standard EBeam conditions the spin-spin interaction is negligible relative to the space charge effect.

## 3. Non-diffracting range and effective current vs. beam width, for different apertures

The following figure presents a quantitative comparison in the performance between our shape-preserving multi-electron wavefunction and multi-electron Bessel and Gaussian beams. Similar to the compassion that was made in Fig. 4 in the paper but with lower beam current of $5\mu A$ (recall the total beam current in Fig. 4 in the paper is $I = 50\mu A$). We present the results for two apertures: $140nm$ (as in the example in Fig.3) and $420nm$, both under the same acceleration voltage of $20kV$. The aperture determines the wavefunction density because of normalization requirement from Eq. (iv). Therefore the aperture size affects the nonlinear solution and the non-diffracting range of the EBeam, even for a fixed beam current.

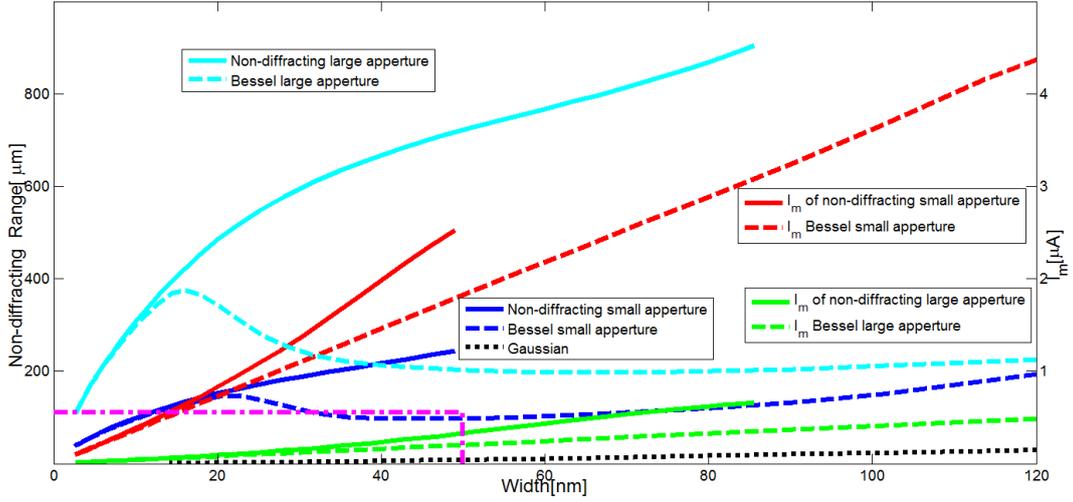

We examine the non-diffracting range for the three types of initial wavefunctions, plotting our shape-preserving wavefunction in solid lines, the Bessel beam in dashed lines, and the Gaussian beam in black dotted line. We compare the case of wide aperture (cyan curves) and small aperture (blue curves). The current inside the main lobe of the beam is denoted in green for the large aperture and in red for the small aperture. We can see tendencies similar to the results presented in Fig. 4. However, the Critical width here is much larger, showing strong dependence on the aperture, while only weak dependence on the total current. Additionally, as the aperture is increased, the EBeam density decreases (because of total probability normalization) and the EBeam experiences less repulsion, resulting in larger non-diffracting range. This explains the larger non-diffracting range of the wide aperture when compare to the smaller aperture (see cyan vs. blue curves), but also leads to the lower current inside the main lobe (green vs. red). Further comparison is shown in Fig. 4 in the main text, which is the zoom-in of the area marked by the magenta rectangle.